\documentclass[sn-mathphys]{sn-jnl}

\jyear{2021}%

\theoremstyle{thmstyleone}%
\theoremstyle{thmstyletwo}%

\theoremstyle{thmstylethree}%

\usepackage{tabularx}
\usepackage{adjustbox}
\usepackage{subfig}
\usepackage{amsmath}
\usepackage{booktabs}
\usepackage{multirow}
\usepackage{mathrsfs} 
\usepackage{amssymb}
\usepackage{float}
\usepackage{amssymb}
\usepackage{float}
\usepackage{lipsum}
\usepackage{ifmtarg}
\usepackage{tabularx}
\usepackage{adjustbox}
\usepackage{epstopdf}
\usepackage{caption}
\usepackage{booktabs}
\usepackage{multirow}
\usepackage{subfloat}
\usepackage{caption}
\usepackage{cleveref} 
\usepackage{enumerate}

\begin{document}

\title[ADC-cycleGAN]{{\color{black}An Attentive-based Generative Model for Medical Image Synthesis}}


\author[1]{\fnm{Jiayuan} \sur{Wang}}\email{wang621@uwindsor.ca}

\author*[1]{\fnm{Q. M. Jonathan} \sur{Wu}}\email{jwu@uwindsor.ca}

\author[2]{\fnm{Farhad} \sur{Pourpanah}}\email{farhad.086@gmail.com}

\affil[1]{\orgdiv{ Centre for Computer Vision
and Deep Learning, Department of Electrical and Computer Engineering, University of Windsor, Canada}}
\affil[2]{\orgdiv{Department of Electrical and Computer Engineering, Queens University, Canada}}


\abstract{{\color{black}Magnetic resonance (MR) and computer tomography (CT) imaging are valuable tools for diagnosing diseases and planning treatment. However, limitations such as radiation exposure and cost can restrict access to certain imaging modalities. To address this issue, medical image synthesis can generate one modality from another, but many existing models struggle with high-quality image synthesis when multiple slices are present in the dataset. This study proposes an attention-based dual contrast generative model, called ADC-cycleGAN, which can synthesize medical images from unpaired data with multiple slices. The model integrates a dual contrast loss term with the CycleGAN loss to ensure that the synthesized images are distinguishable from the source domain. Additionally, an attention mechanism is incorporated into the generators to extract informative features from both channel and spatial domains. To improve performance when dealing with multiple slices, the $K$-means algorithm is used to cluster the dataset into $K$ groups, and each group is used to train a separate ADC-cycleGAN. Experimental results demonstrate that the proposed ADC-cycleGAN model produces comparable samples to other state-of-the-art generative models, achieving the highest PSNR and SSIM values of 19.04385 and 0.68551, respectively. 
We publish the code at \textcolor{blue}{https://github.com/JiayuanWang-JW/ADC-cycleGAN}.}}




\maketitle

\section{Introduction}
\label{Sec:intro}
Medical image analysis plays an important role in many clinical applications. 
Two types of neuroimaging techniques, including magnetic resonance (MR) and computer tomography (CT), are being widely used to diagnose various diseases and treatment planning. These modalities provide mutually-complementary information. MR images provide excellent soft tissue contrast and have no ionizing radiation, while CT images are suitable for bony structure and chest analysis~\cite{xu2020bpgan,yang2021synthesizing}. Medical image acquisition is a crucial step for medical image analysis. However, obtaining both modalities is a challenging issue due to multiple factors such as the unavailability of certain modalities, radiation dose, or high cost. Therefore, it is important to develop medical image synthesis models to generate one modality from another~\cite{chen2019one,lee2020spine}.\par

Medical image synthesis methods learn a model to transfer knowledge from one modality, i.e., source domain, into another, i.e., target domain, without utilizing extra annotations from the target domain~\cite{tomar2021self}. In other words, it learns a mapping function to map images from the source domain to the target domain. The learning can be done in a supervised manner using paired data~\cite{merida2015evaluation}, i.e., pairs of MR and CT images belong to the same patient and are perfectly registered, or unsupervised manner using unpaired data~\cite{lian2022cocyclereg,li2020self}. 
Since CT and MR images have different structures, learning a direct mapping function between them is a challenging issue. Thus, the mapping function has to be complex and highly non-linear to bridge the structural differences between the two modalities.\par

Early medical image synthesis methods are based on segmentation and atlas~\cite{jiao2020self}. 
Segmentation methods, first, segment an image from the source domain into several tissue classes and then synthesize the corresponding image in the target domain by intensity-filling of each class~\cite{berker2012mri}. 
In contrast, atlas methods, first, register each image from the source domain into its corresponding atlas via a transformation and then apply the registration to the target domain atlas to synthesize the corresponding sample in the target domain~\cite{sjolund2015generating}. The quality of the synthesized images by these methods relies on the segmentation and atlas quality.

{\color{black}Recently, convolutional neural networks (CNNs) methods have shown remarkable results in medical areas, such as disease diagnosis~\cite{bhosale2022application,bhosale2023puldi}, especially in synthesizing medical image tasks due to their ability in extracting task-specific features.} For example, Li et al.~\cite{li2014deep} developed a CNNs-based model to use MR images and generate the corresponding positron emission tomography (PET) image for the same subject.
Huang et al.~\cite{huang2017simultaneous} proposed a weakly-supervised joint convolutional sparse coding model to simultaneously conduct super-resolution and medical image synthesis tasks. Zhao et al.~\cite{zhao2018towards} designed a CNN architecture to learn a mapping between two modalities utilizing imperfect registered CT-MR pairs.
\par


Generative adversarial networks (GANs)~\cite{goodfellow2014generative} methods have produced promising results in synthesizing medical images. Nie et al.~\cite{nie2018medical} integrated adversarial training strategy into a fully convolutional network (FCN) to model non-linear mapping between two modalities. 
Dalmaz et al.~\cite{dalmaz2022resvit} introduced a novel GAN-based method that leverages the contextual sensitivity of vision transformers, the precision of convolutional operators, and the realism of adversarial learning to improve image generation.
However, these generative models need a large number of perfectly registered paired data, which is a challenging issue since paired data are typically scarce.
To alleviate the paired data restriction, many studies adopted CycleGAN~\cite{zhu2017unpaired} structure to convert medical image synthesis task into image-to-image translation that can learn from unpaired data. However, CycleGAN cannot perform well in transferring complex texture domain such as CT-MR, and it needs additional term(s) to learn a better mapping function between two modalities and consequently enhance the quality of the synthesized images in the target domain.
For example, in~\cite{liu2021unpaired}, an adversarial learning model for synthesizing Ki-67-stained images from H\&E-stained images from unpaired data has been proposed. This model attempts to preserve the structural details of the synthesized images by adopting the structural similarity constraint and skip connection.  
Huo et al.~\cite{huo2018synseg} introduced an end-to-end synthetic segmentation model to perform segmentation in the target domain without having access to any manual labels.
Chen et al.~\cite{chen2019one} presented a one-shot generative model for MR image segmentation that utilizes unpaired data in addition to a single paired CT-MR dataset. This model consists of two networks, which are cross-modality image synthesis and MR image segmentation, that are jointly trained.\par

Moreover, several studies~\cite{liu2020cbct,huang2020cagan,xu2019semi,nie2020adversarial,tomar2021self,Heran2020Unsupervised} integrated attention mechanisms into the model to focus on the most important regions of the image.  
Studies~\cite{liu2020cbct,huang2020cagan,xu2019semi} integrated attention mechanism into a conditional GAN to expand the receptive field and extract richer contextual dependencies.
In~\cite{nie2020adversarial}, a difficulty-aware attention mechanism that considers the structural information in order to handle hard samples or regions. 
Tomar et al.~\cite{tomar2021self} introduced a self-attention mechanism for attending various structures of the organ by leveraging an auxiliary semantic segmentation information. Yang et al.~\cite{Heran2020Unsupervised} incorporated self-attention mechanism into the generators for modelling long-range spatial dependencies in the synthesized images.\par

In our previous work~\cite{wang2022dc}, we proposed DC-cycleGAN, which is a bidirectional generative model, for medical image synthesis. We introduced a new loss term, called dual contrast (DC) loss, to enhance the model performance. DC loss locates the synthesized images far away from the samples of the source domain. To accomplish this, the DC loss uses the samples from the source domain as negative samples. 
We evaluated the performance of our model using 100 samples selected from a dataset proposed by Han et al.~\cite{han2017mr}. 
However, DC-cycleGAN and other methods produce unstable results as this dataset contains various slices. Abu et al.~\cite{abu2021paired} alleviated this problem using auxiliary samples.\par

{\color{black}
In this paper, we propose an \textit{attention-based dual contrast CycleGAN} (ADC-cycleGAN) to further improve the performance of the DC-cycleGAN model and address the above-mentioned limitation of existing methods. The main contributions of our study are as follows:
\begin{enumerate}
    \item Convolutional block attention module (CBAM)~\cite{woo2018cbam} is integrated into the DC-cycleGAN structure, namely ADC-cycleGAN, to extract more informative features from both channel and space dimensions in synthesizing medical images. 
    \item To generate high-quality images from datasets with multiple structures. To achieve this, the $K$-means algorithm is employed to cluster the training dataset into $K$ groups and then each group is used to train an ADC-cycleGAN model, i.e., $K$ models are trained. Using the clustering algorithm reduces the complexity of the dataset and alleviates the generator collapse.
    \item To evaluate the performance of our proposed bidirectional medical image synthesis method with baseline and other state-of-the-art methods.
\end{enumerate}

}


This paper contains five sections. Section~\ref{Sec:related} reviews the existing methods. Section~\ref{Sec:model} presents the proposed model. Section~\ref{Sec:exp} provides the experimental results and ablation studies. Finally,  the concluding remarks and future research directions are presented in Section~\ref{Sec:con}.\par

\section{Related works}
\label{Sec:related}
This section reviews recent advances in medical image synthesis and attention mechanism.\par

\subsection{Medical image synthesis}
\label{Sec:sex:medical}
Medical image synthesis is an active area of research. On one hand, it aims to reduce time, labor, and cost \cite{wang2021review}. On the other hand, some patients have metal devices in their bodies in which they can not scan the MR images. Medical image synthesis techniques can be broadly categorized into traditional- and deep learning (DL)-based methods. Traditional methods learn a mapping function between similar patches from the two domains. This category can be grouped into atlas-~\cite{hofmann2008mri,chen2015using,dowling2012atlas} and segmentation-~\cite{berker2012mri,izquierdo2014spm8,delpon2016comparison,hsu2013investigation} methods.
Izquierdo et al.~\cite{izquierdo2014spm8} first segmented the MR images into 6 tissue classes, and then uses a diffeomorphic approach to non-rigidly co-register. In the same way, the anatomical MR data for new subjects is co-registered with the template. Finally, the inverse transformations were applied to synthesize CT scans. Burgos et al.~\cite{burgos2014attenuation} generated CT scans and attenuation maps to enhance the attenuation correction for PET/MR
scanners. As such, CT scans are synthesized using a multi-atlas information propagation scheme, in which a local image similarity measure is used to locally match the MRI-derived patient’s morphology to a database of MRI/CT pairs.\par

DL methods can be classified into Auto-encoder(AE), GAN, and U-net~\cite{wang2021review}. 
DEDIS~\cite{sevetlidis2016whole}, which is a deep encoder-decoder image synthesizer, performs MR image translation into different modalities, i.e., synthesizes T2 from T1 and DWI from T2. DEDIS is fast, requires a lower computational cost, and can produce comparable results as compared with the traditional methods.  
Hi-Net~\cite{zhou2020hi} solves the missing modality problem in medical imaging by learning a mapping function from the multi-modal domains to the target domain. It consists of three components, including a modality-specific network that learns the features of each modality, a multi-modal fusion network that learns common latent features of multi-modal data, and a multi-modal synthesis network that combines the learned latent features with hierarchical features from each modality to synthesize target images.  
Auto-GAN~\cite{cao2020auto} is a self-supervised AE structure that obtains target-modality-specific information for the generator to synthesize missing MR image modality from available modalities.  
SkrGAN~\cite{zhang2019skrgan} integrates a sketch prior constraint into the GAN to synthesize high-quality medical images. It also embeds a sketch-based representation using a color render mapping technique. In another study~\cite{hu2019cross}, U-Net is combined with an adversarial training strategy to synthesize 2D PET from  MR slices. \par

A number of models based on CycleGAN have been introduced for medical image synthesis from unpaired data. For example, UC-GAN~\cite{wu2019uc} integrates U-Net into the CycleGAN structure for synthesizing CT scans from MR images. 
SC-cycleGAN~\cite{yang2020unsupervised} integrates structure-consistency loss with spectral normalization and self-attention mechanism for generating CT from MR images. 
{\color{black}Nice-GAN~\cite{chen2020reusing} utilizes the discriminator's encoding capability to improve the quality of generated images in the target domain. To ensure a stable training process during the adversarial min-max game, a decoupled training strategy is developed. This strategy prevents any training inconsistency and enables the encoder to train effectively by maximizing the loss and keeping it frozen otherwise.
  }

In~\cite{lee2021unsupervised}, the CBAM is integrated into $\beta$-cycleGAN to focus on the most important channel and spatial features.
{\color{black} RegGAN~\cite{kong2021breaking} is based on the theory of "loss-correction" that is proposed for translating MR T1 to T2. Although RegGAN can produce promising results, it is a single-direction synthesis model. 
UGATIT~\cite{citation-0} is an unsupervised method that integrates an attention module and a normalization function into its structure. }

{\color{green} 

}

\subsection{Attention}
\label{Sec:sex:att}
The role of attention in human perception is crucial. The human visual system focuses on the most salient parts of an image to explore the visual structure instead of processing the whole scene at once~\cite{larochelle2010learning}. Inspired by the human visual system, attention processing has been incorporated into DL to enhance the model's performance~\cite{fukui2019attention,fu2019dual}.
The attention mechanism is a crucial component in computer vision (CV), applied in various application domains such as text classification~\cite{liu2019bidirectional}, machine translation~\cite{vaswani2017attention} and image classification~\cite{chen2021crossvit}, just to name a few.\par

The attention mechanism in CV can be broadly divided into four categories: channel attention to focus on ``what," spatial attention to focus on ``where," temporal attention to focus on ``when," and branch attention to focus on ``which"~\cite{guo2022attention}. Meantime, several studies have explored the combinations of two or more of them to enhance the model performance, e.g., spatial \& temporal, and channel \& spatial. 
The channel \& spatial attention simultaneously take advantage of both channel and spatial attention to focus on important objects and regions of the images. The representative models of this category include CBAM~\cite{woo2018cbam}, dual attention~\cite{fu2019dual}, and triplet attention~\cite{misra2021rotate}. 
Among them, CBAM is a simple and effective model that considers both channel and spatial domains. It can be integrated into CNN-based architectures to perform end-to-end training with negligible overhead~\cite{wang2021avnc}. Therefore, we integrated CBAM into our generators to extract features from both channel and space domains. We discuss the CBAM structure in Section~\ref{attention mechanism}.

\section{Proposed model}
\label{Sec:model}
This section introduces ADC-cycleGAN, a method for synthesizing medical images from unpaired data.
We begin by formulating the problem and providing an overview of our approach. Next, we present specific instantiations of our method. {\color{black} To aid the reader, we list common symbols along with their corresponding descriptions in Table~\ref{symbols}.

\begin{table}[h]
\centering
\vspace{-0.5cm}
\color{black} 
\caption{Table of common symbols.}
\begin{adjustbox} {width=\columnwidth}
\label{symbols}
    \begin{tabular}{c c}
        \toprule Symbol & \multicolumn{1}{c}{ Description } \\
        \midrule x & real CT image sample \\
        \midrule y & real MR image sample \\
        \midrule G(x) \& $\hat{y}$ & synthesized MR images \\
        \midrule F(y) \& $\hat{x}$ & synthesized CT images \\
        \midrule D(x) & discriminator to identify real CT images from the synthetic or randomly selected image from the source domain\\
        \midrule D(y) & discriminator to identify real MR images from the synthetic or randomly selected image from the source domain \\
        \midrule $\lambda$ & weight of cycle consistency loss in the whole loss functions \\
        \midrule $\beta$ & weight of dual contrast loss in the whole loss functions \\
        \bottomrule
        \end{tabular}
\end{adjustbox}
\end{table}

}

\subsection{Problem formulation}
\label{Sec:sec:prob}
Assume $X=\{x_i\}_{i=1}^N$ indicates the set of CT images and $Y\{y_j\}_{j=1}^M$ represents the set of MR images. 

The objective is to train a bidirectional projection function between CT and MR images using unpaired data. To accomplish this, ADC-cycleGAN employs two generators: $G:x\to \hat{y}$ and $F:y\to \hat{x}$, to learn the CT-to-MR and MR-to-CT mappings, respectively. Here, $\hat{y}=G(x)$ and $\hat{x}=F(y)$ denote the synthesized MR and CT images, respectively. Additionally, two discriminators $D_Y$ and $D_X$ are used to differentiate between real MR and CT images and synthetic or randomly selected images from the source domain.\par


\subsection{Model overview}
\label{Sec:sec:model}
\begin{figure}[h]
\centering
\includegraphics[width=0.99\linewidth]{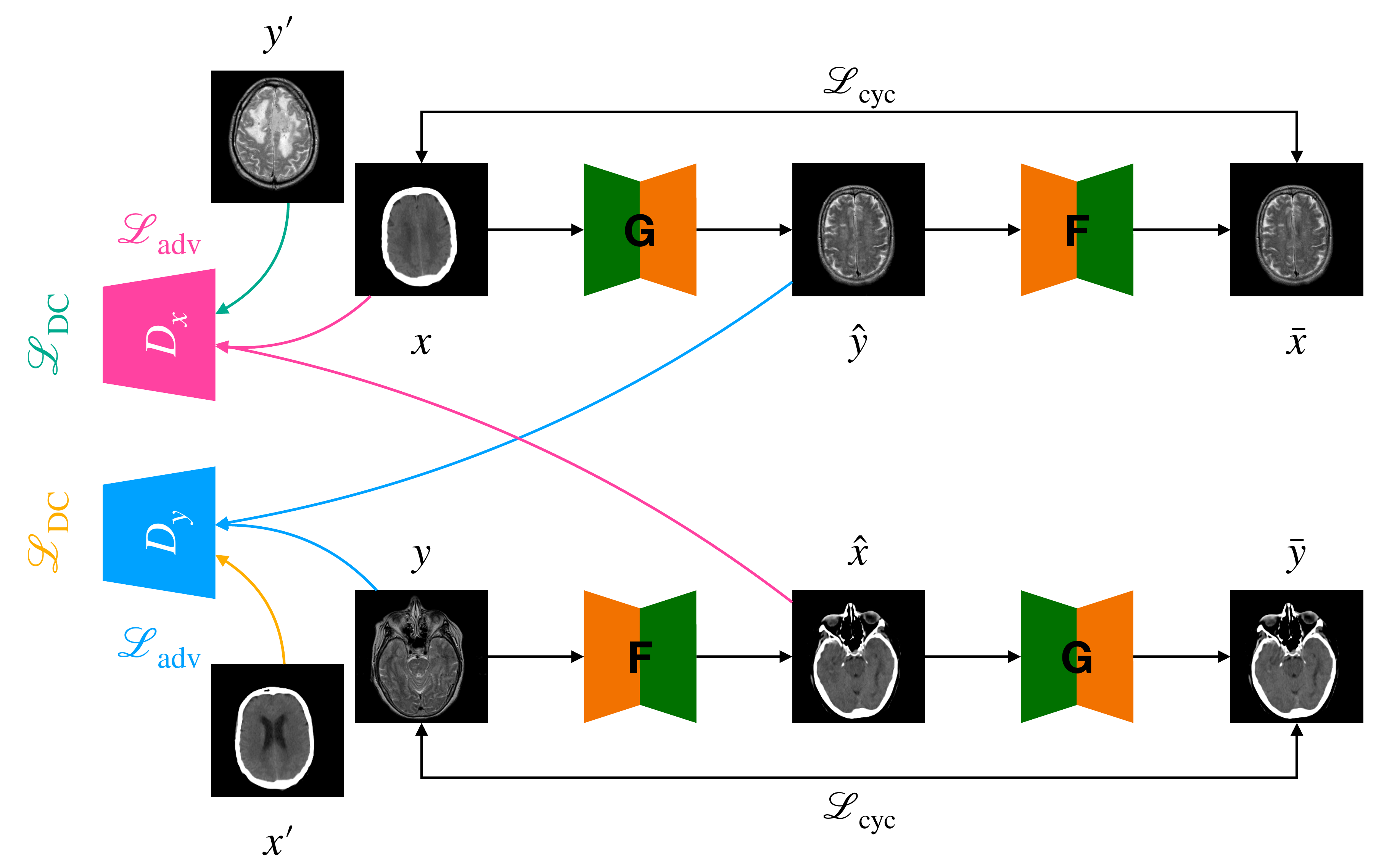}
\caption{The structure of ADC-cycleGAN}
\label{ADC-cyclegan}
\end{figure}
First, the training set is clustered into a number of groups using $K$-means algorithm. Then, a model, i.e., ADC-cycleGAN (see Fig.~\ref{ADC-cyclegan}), is trained based on samples of each group.
Fig.~\ref{ADC-cyclegan} illustrates the overview of ADC-cycleGAN. It is a bidirectional generative model that combines CycleGAN with dual contrast loss~\cite{wang2022dc} to synthesize high-quality images in the target domain.
Each component of ADC-cycleGAN is discussed in detail as follows.\par

\subsubsection{CycleGAN}
\label{Sec:sec:GAN}

CycleGAN, which is originally proposed by Zhu et al.~\cite{zhu2017unpaired} for unpaired image-to-image translation, consists of a GAN~\cite{goodfellow2014generative} and cycle consistency loss.
The GAN is composed of a generator $G:X\to Y$ to synthesize images in the target domain by transferring knowledge from the source domain, and a discriminator $D_Y$ to identify real images in the target domain from the synthesized ones.
These modules play a two-player game in which the generator forces the discriminator to enhance its distinguishing ability, while the discriminator forces the generator to synthesize more realistic images.
In another word, the generator learns a mapping between source and target domains $X\to Y$, and the discriminator's output is a probability that indicates the input image is real.  
The generator and discriminator play a min-max game to optimize using adversarial loss function as:
\begin{multline}\label{CycleGAN_adversarial_loss}
\mathcal{L}_{\mathrm{GAN}}\left(G, D_{Y}, X, Y\right) =\mathbb{E}_{y \sim p_{\mathrm{data}}(y)}\left[\log D_{Y}(y)\right] \\
+\mathbb{E}_{x \sim p_{\mathrm{data}}(x)}\left[\log \left(1-D_{Y}(G(x))\right)]\right..
\end{multline}
where $x$ and $y$ are real images from the source and target domains, respectively, $G(x)$ is the synthesized image in the target domain. 

Although adversarial learning can learn a mapping between the source and target domains, it has the limitation of mapping input images to any arbitrary location in the target domain. Consequently, adversarial learning does not guarantee that the learned function can accurately map input $x_i$ to its intended output $y_i$. Furthermore, acquiring paired image datasets can be challenging in practice.
To overcome these issues, it is essential to reconstruct the synthesized images into their target domain via a cycle consistency loss. To accomplish this, an additional generator $F:Y\to X$ is necessary for reconstructing real source domain images. A discriminator $D_X$ is also required to distinguish real images from the synthesized ones in the target domain

\begin{multline}\label{cycle_consistency_loss}
\mathcal{L}_{\mathrm{cycle}}(G, F) =\mathbb{E}_{x \sim p_{\text {data }}(x)}\left[\|F(G(x))-x\|_{1}\right] \\
+\mathbb{E}_{y \sim p_{\text {data }}(y)}\left[\|G(F(y))-y\|_{1}\right],
\end{multline}
where $F(G(x))$ and $G(F(y))$ indicate the reconstructed CT and MR images, respectively.\par

To summarize, the CycleGAN is comprised of four distinct networks: two generators ($G$ and $F$) and two discriminators ($D_{X}$ and $D_{Y}$). 
The object function for the CycleGAN is as follows:
\begin{multline}\label{adv}
\mathcal{L}\left(G, F, D_{X}, D_{Y}\right)=\mathcal{L}_{\mathrm{GAN}}\left(G, D_{Y}, X, Y\right) \\
+\mathcal{L}_{\mathrm{GAN}}\left(F, D_{X}, Y, X\right)
+\lambda \mathcal{L}_{\mathrm{cycle}}(G, F),
\end{multline}
where $\lambda$ indicates the weight of cycle consistency loss.\par

The aim to solve is:
\begin{align}
G^{*}, F^{*}= \arg \min _{G, F} \max _{D_{X}, D_{Y}} \mathcal{L}\left(G, F, D_{X}, D_{Y}\right).
\end{align}

\subsubsection{Dual contrast}
\label{Sec:sex:DC}
Since CT and MR images have different structures and there is no constraints between real source and synthesized images~\cite{Heran2020Unsupervised}, CycleGAN alone can not synthesize high-quality images in the target domain. 
To mitigate this issue, an additional loss term, known as \textit{dual contrast} (DC) loss, is introduced into discriminators. The DC loss utilizes samples from the source domain as negative samples ($x'$ and $y'$) to prompt the model to generate images that are distinct from the source domain (refer to Fig.~\ref{ADC-cyclegan}).  Instead of discriminating between real images ($y$ and $x$) and synthesized images ($\hat{y}$ and $\hat{x}$), the discriminators are tasked with distinguishing real images in the target domain from both synthesized images and randomly selected images from the source domain. 
In other words, the real images are assigned to class 1, whereas synthesized images and randomly selected samples from the source domain are assigned to class 0.
The dual contrast loss function, as follows:
\begin{align}
\mathcal{L}_{\mathrm{DC}}\left(D_{Y}, X, Y\right)= \mathbb{E}_{x' \sim p_{\mathrm{data}}(x')}\left[\log (1-D_{Y}\left(x'\right))\right].
\end{align}

\subsubsection{Overall objective function}
\label{Sec:sex:ob}
By adding DC loss into (\ref{adv}), the final objective function can be written as:
\begin{multline}
\mathcal{L}\left(G, F, D_{X}, D_{Y}\right)= \mathcal{L}_{\mathrm{GAN}}\left(G, D_{Y}, X, Y\right)+\\ \mathcal{L}_{\mathrm{GAN}}\left(F, D_{X}, Y, X\right)
+\beta \mathcal{L}_{\mathrm{DC}}\left(D_{Y}, X, Y\right)+\\ \beta \mathcal{L}_{\mathrm{DC}}\left(D_{X}, Y, X\right)
+\lambda \mathcal{L}_{\mathrm{cycle}}(G, F),
\end{multline}

where $\lambda$ and $\beta$ indicate the weight of cycle consistency loss and dual contrast loss in the whole loss functions, respectively.

\subsection{Attention mechanism}
\label{attention mechanism}
\begin{figure}[h]
        \centering
        \subfloat{
		\begin{minipage}[t]{0.8\textwidth}
			\centering

			\includegraphics[width=0.95\textwidth]{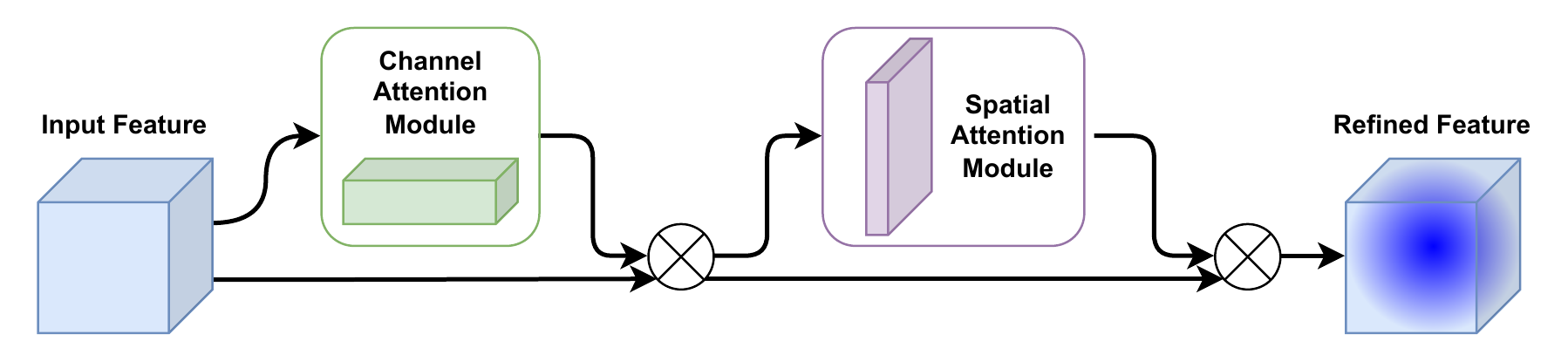}
			\centerline{(a) Convolutional Block Attention Module}
			\vspace{0.1cm}
			\includegraphics[width=0.95\textwidth]{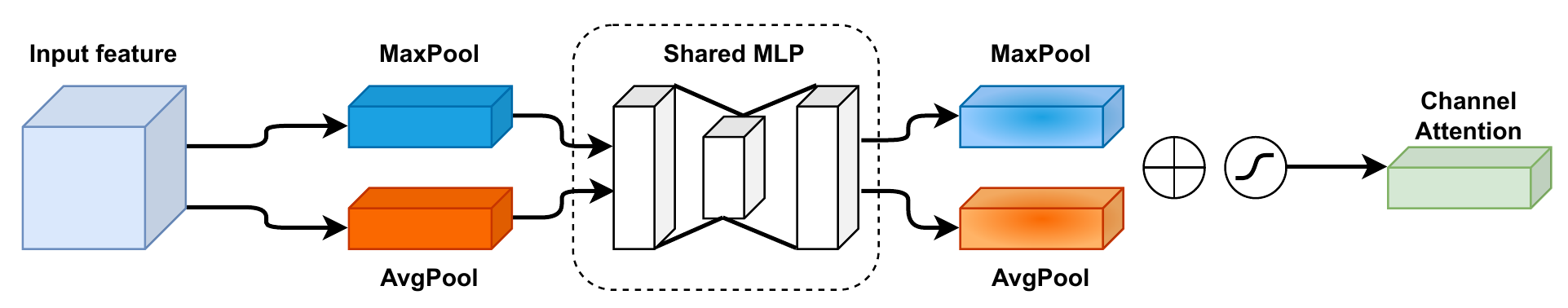}
			\centerline{(b) Channel Attention Module}
			\vspace{0.1cm}
			\includegraphics[width=0.95\textwidth]{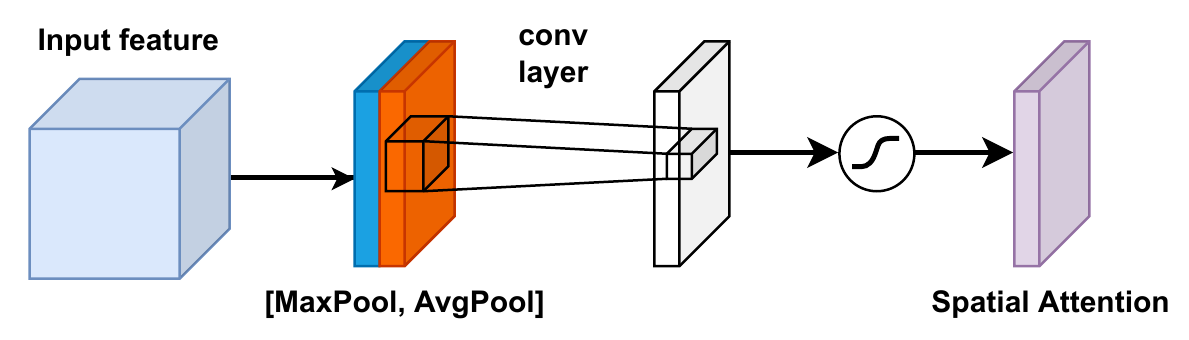}
			\centerline{(c) Spatial Attention Module}
		\end{minipage}%
	}%

        \caption{Structure of CBAM and sub modules}
        \label{CBAM_structure}
\end{figure}
Channel attention, spatial attention, and their combinations, i.e., mixed attention, are often used in computer vision tasks. Channel attention focuses on \textit{``what'' are important channels} to pay attention in a feature map. It adds a different weight to each channel, and the weight with a high value is more correlated. 
In contrast, spatial attention focuses on \textit{``where'' is an informative part} to pay attention by learning a weight on a 2D feature map. 
While, mixed attention combines channel and spatial attentions. CBAM~\cite{woo2018cbam} is the most representative model of this category that has been widely integrated into the CNN-based models to improve performance with negligible overheads. It is presented in detail, as follows.\par

Assume $\mathbb{R}^{C \times H \times W}$ represents an input feature map, where $C$, $H$ and $W$ indicate channel, high and wide, respectively.
As shown in Fig.~\ref{CBAM_structure} (a), CBAM consists of two sub-modules, including channel and spatial, that sequentially infers a 1D channel attention map $\mathbf{M}_{\mathbf{c}} \in \mathbb{R}^{C \times 1 \times 1}$ and a 2D spatial attention map $\mathbf{M}_{\mathbf{s}} \in \mathbb{R}^{1 \times H \times W}$. The CBAM whole process can be expressed as:
\begin{equation}
\begin{aligned}
\mathbf{F}^{\prime} &=\mathbf{M}_{\mathbf{c}}(\mathbf{F}) \otimes \mathbf{F}, \\
\mathbf{F}^{\prime \prime} &=\mathbf{M}_{\mathbf{s}}\left(\mathbf{F}^{\prime}\right) \otimes \mathbf{F}^{\prime},
\end{aligned}
\end{equation}
where $\otimes$ is the element-wise multiplication, $\mathbf{F}^{\prime}$ indicates a feature map after combining the input feature $\mathbf{F}$ and channel attention map $\mathbf{M}_{\mathbf{c}}(\mathbf{F})$, and $\mathbf{F}^{\prime \prime}$ is the final refined feature.\par

It first uses max-pooling and average-pooling operations to produce two spatial context descriptors $\mathbf{F}_{\mathbf{a v g}}^{\mathrm{c}}$ and $\mathbf{F}_{\mathbf{m a x}}^{\mathrm{c}}$, and then, the channel attention map $\mathbf{M}_{\mathbf{c}} \in \mathbb{R}^{C \times 1 \times 1}$ is produced by feeding the descriptors into a shared network, i.e., an MLP with one hidden layer, as: 
\begin{multline}
\mathbf{M}_{\mathbf{c}}(\mathbf{F}) =\sigma(M L P(\operatorname{AvgPool}(\mathbf{F}))+M L P(\operatorname{MaxPool}(\mathbf{F}))) \\
=\sigma\left(\mathbf{W}_{\mathbf{1}}\left(\mathbf{W}_{\mathbf{0}}\left(\mathbf{F}_{\mathbf{a v g}}^{\mathbf{c}}\right)\right)+\mathbf{W}_{\mathbf{1}}\left(\mathbf{W}_{\mathbf{0}}\left(\mathbf{F}_{\mathbf{m a x }}^{\mathbf{c}}\right)\right)\right),
\end{multline}
where $\sigma$ is sigmoid activation function, and $\mathbf{W}_{\mathbf{0}} \in \mathbb{R}^{C / r \times C}$ and $\mathbf{W}_{\mathbf{1}} \in \mathbb{R}^{C \times C / r}$ are the weights of the shared network, where $r$ indicates the reduction ratio which is set to 8 in this study.\par

\textbf{Spatial attention module:} Fig.~\ref{CBAM_structure} (c) depicts the spatial attention module, which utilizes the inter-spatial relationship of features to produce the spatial attention map. Unlike the channel attention module, this module applies max-pooling and average-pooling operations to the channel axis, resulting in two 2D maps $\mathbf{F}_{\mathbf{a v g}}^{\mathbf{s}} \in \mathbb{R}^{1 \times H \times W}$ and $\mathbf{F}_{\mathbf{m a x}}^{\mathbf{s}} \in \mathbb{R}^{1 \times H \times W}$. These maps are concatenated to generate a feature descriptor.
Then, a convolution layer is applied to produce a 2D spatial attention map $\mathbf{M}_{\mathbf{s}}(\mathbf{F})$, as follows:  

\begin{equation}
\begin{aligned}
\mathbf{M}_{\mathbf{s}}(\mathbf{F}) &=\sigma\left(f^{7 \times 7}([\operatorname{AvgPool}(\mathbf{F}) ; \operatorname{MaxPool}(\mathbf{F})])\right) \\
&=\sigma\left(f^{7 \times 7}\left(\left[\mathbf{F}_{\mathbf{a v g}}^{\mathbf{s}} ; \mathbf{F}_{\mathbf{m a x}}^{\mathbf{s}}\right]\right)\right)
\end{aligned}
\end{equation}
where $\sigma$ indicates the sigmoid activation functions, $f^{7 \times 7}$ indicate a convolution layer with a 7 x 7 filter.


\subsection{Model structure}
\label{Sec:sex:str}
This subsection discusses the structures of $K$-means algorithm, generators and discriminators in detail.\par

\subsubsection{k-means algorithm}

\begin{figure}[h]
\centering
\includegraphics[width=0.99\linewidth]{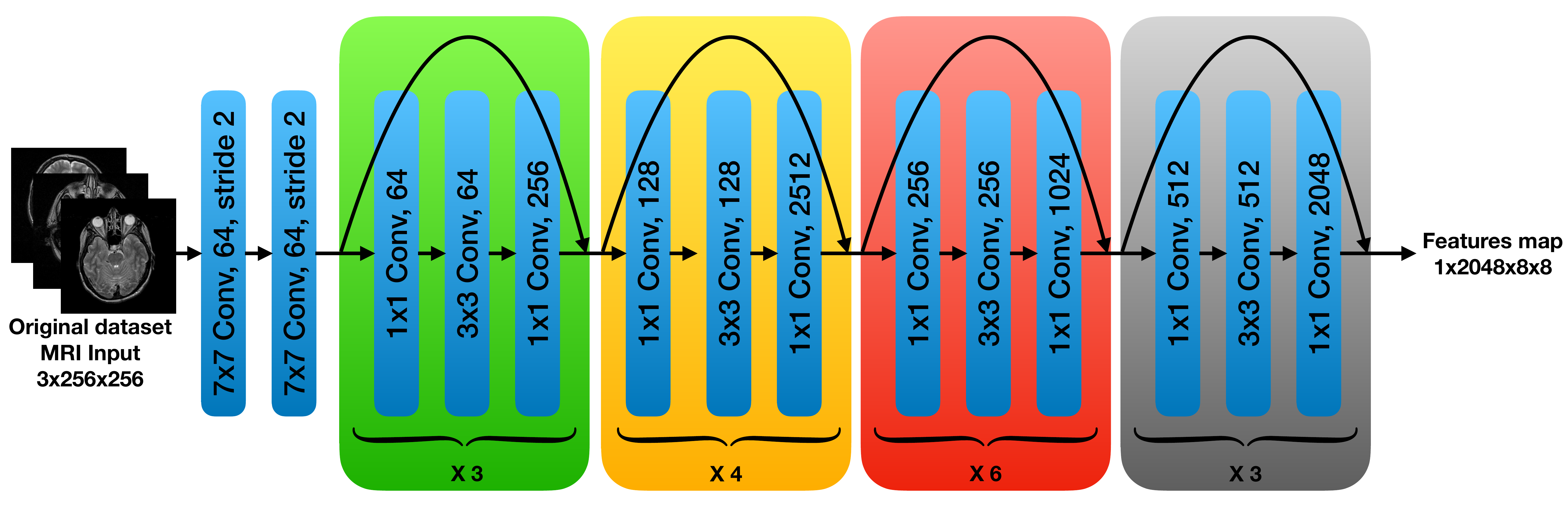}
\caption{The preprocessing of cluster.}
\label{resnet50}
\end{figure}

To cluster the training set, we first use the Resnet50~\cite{he2016deep} to extract features (see Fig.~\ref{resnet50}), and then, the `KMeans' function from the sklearn package is used for clustering. {\color{black} We initialized the cluster centers randomly using the `KMeans' function in the sklearn package. Additionally, we used the $K$-means algorithm with the same clustering results for all methods by clustering the training dataset into $K$ groups. We repeated this procedure five times to ensure the consistency of the results and avoid dependency on the initial random seed.} The KMeans function uses the Euclidean distance to continuously compute the distance between samples and the clusters' center, and update their centroids by optimizing:   
\begin{equation}
E=\sum_{i=1}^K \sum_{\boldsymbol{x} \in C_i}\left\|\boldsymbol{x}-\boldsymbol{\mu}_i\right\|_2^2,
\end{equation}
where $K$ is the number of clusters, $\boldsymbol{\mu}_i$ represents the center of the $i$-th cluster, and $\boldsymbol{x} \in C_i$ means $x$ belong to the $i$-th cluster.\par 

\subsubsection{Generator}
\begin{figure*}[h]
\centering
\includegraphics[width=0.99\linewidth]{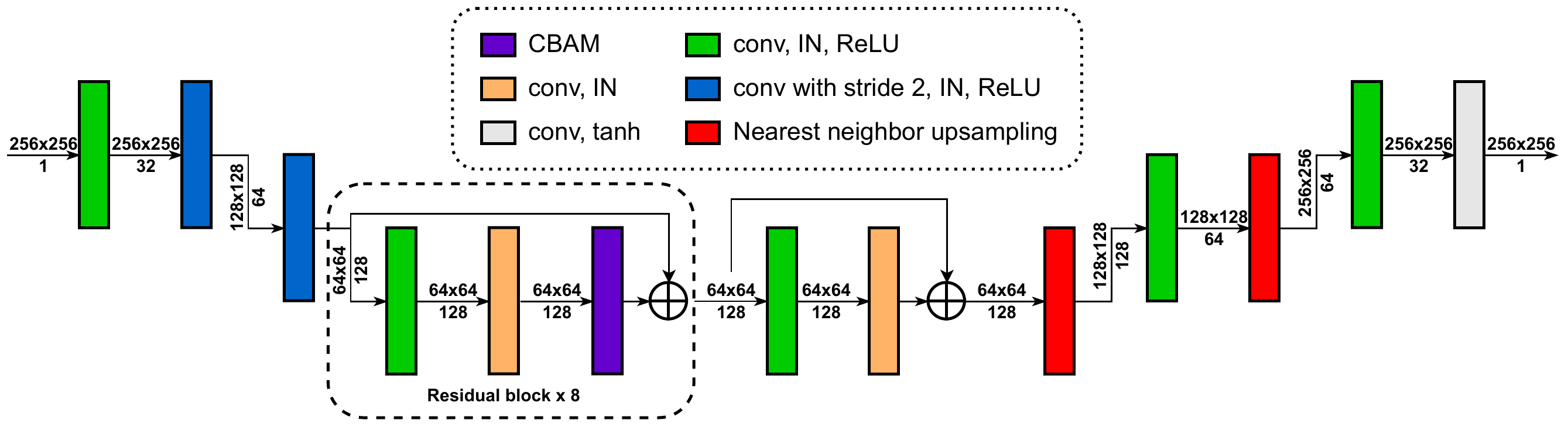}
\caption{The structure of generator integrated into the ADC-cycleGAN model.}
\label{generators}
\end{figure*}
In this study, CBAM~\cite{woo2018cbam} is integrated into the generators to extract more informative features from channel and space domains (see Fig.~\ref{generators}). 
Each generator has one instance normalization (IN) convolutional layer with ReLU activation function, two stride-2 IN-convolutional layers with ReLU activation function, nine residual blocks, in which CBAM is added into the first eight blocks, two blocks of nearest neighbor up-sampling and IN-convolutional layers with ReLU activation function, followed by a convolutional layer with tanh activation function as output layer.\par 

The vanilla CycleGAN utilizes mean absolute error (MAE) loss in generators. However, using MAE may result in synthesizing low-quality images that can not be recognized by human~\cite{snell2017learning,wang2022dc}. To alleviate this issue, Snell et al.~\cite{snell2017learning} showed that using SSIM instead of MAE improves the quality of the reconstructed images. This is mainly due to the considering luminance \emph{l}, contrast \emph{c}, and structure \emph{s}, as follows:
\begin{equation}\label{SSIM}
\operatorname{SSIM}(x_{1}, x_{2})=\frac{\left(2 \mu_{x_{1}} \mu_{x_{2}}+c_{1}\right)\left(2 \sigma_{x_{1} x_{2}}+c_{2}\right)}{\left(\mu_{x_{1}}^{2}+\mu_{x_{2}}^{2}+c_{1}\right)\left(\sigma_{x_{1}}^{2}+\sigma_{x_{2}}^{2}+c_{2}\right)},
\end{equation}
where $c_{1}$ and $c_{2}$ and $c_{3}=\frac{c_{2}}{2}$ are constant values, $\mu_{x_{i}}$ is the mean of the $i$-th image ($i=1,~2$), $\sigma_{x_{i}}$ is th standard deviation of the $i$-th image ($i=1,~2$), and $\sigma_{x_{1} x_{2}}$ represents the covariance of $x_{1}$ and $x_{2}$. while: 
\begin{equation}\label{formulal}
l(x_{1}, x_{2})=\frac{2 \mu_{x_{1}} \mu_{x_{2}}+c_{1}}{\mu_{x_{1}}^{2}+\mu_{x_{2}}^{2}+c_{1}},
\end{equation}
\begin{equation}\label{formulac}
c(x_{1}, x_{2})=\frac{2 \sigma_{x_{1}} \sigma_{x_{2}}+c_{2}}{\sigma_{x_{1}}^{2}+\sigma_{x_{2}}^{2}+c_{2}},
\end{equation}
\begin{equation}\label{formulas}
s(x_{1}, x_{2})=\frac{\sigma_{x_{1} x_{2}}+c_{3}}{\sigma_{x_{1}} \sigma_{x_{2}}+c_{3}}.
\end{equation}

Therefore, in this study, we use SSIM instead of MAE in the cycle consistency loss. The Eq.~(\ref{cycle_consistency_loss}) can be written as:
\begin{equation}\label{scycle_consistency_loss}
\mathcal{L}_{\mathrm{cycle}}(G, F) =(1-SSIM(F(G(x)), x))
+(1-SSIM(G(F(y)), y)).
\end{equation}

\subsubsection{Discriminator}
\begin{figure}[h]
\centering
\includegraphics[width=0.99\linewidth]{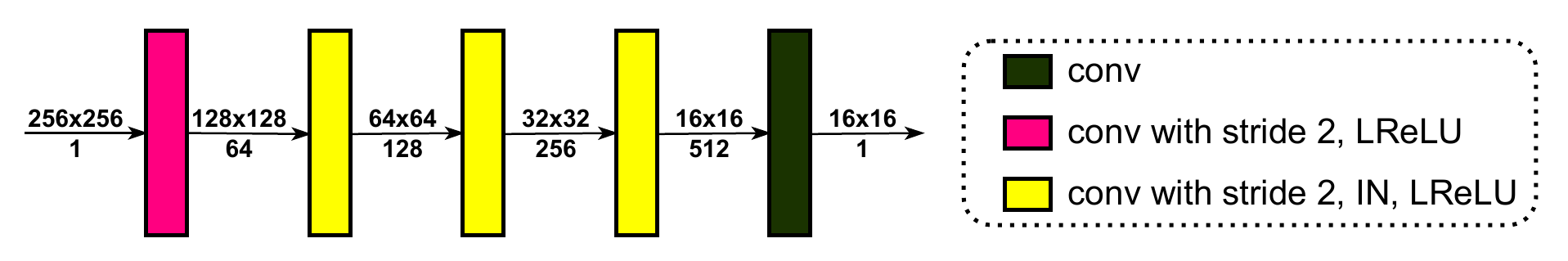}
\caption{The structure of discriminator integrated into the ADC-cycleGAN model.}
\label{discriminators}
\end{figure}
Fig.~\ref{discriminators} shows the structure of discriminators integrated into the ADC-cycleGAN structure. Each discriminator has a stride-2 convolution layer and three stride-2 IN-convolutions layers with LReLU activation function, and the output layer is 94$\times$94 overlapping patches for identifying Whether the input image belongs to class 1 (real) or class 0 (synthesized or a random image from the source domain).\par

Unlike the conventional CycleGAN that uses mean squared error (MSE) loss in the discriminators, in this study, we use cross entropy (CE) because CE converges fast with a small back-propagation error as compared with MSE~\cite{zhou2019mpce}. On one hand, MSE suffers from the problem of gradient vanishing problem in the output layer of networks~\cite{zhong2020generative}. 
In this study, we employed binary CE (BCE) loss at the output layer of the discriminators to differentiate between input images belonging to class 1 (real) and class 0 (synthesized or from the source domain). The BCE is defined as follows:
\begin{equation}\label{CE}
C E(t, y)=-\frac{1}{N} \sum_{i=1}^{N} t_{i} \cdot \log \left(y_{i}\right)+\left(1-t_{i}\right) \cdot \log \left(1-y_{i}\right),
\end{equation}
where $N=256$ due to our discriminator output has 16$\times$16 patches, $t_{i}$ is a label for $i$\textit{th} patch and $y_{i}$ is the predicted probability for $i$-\textit{th} patch.\par

\subsection{Summary of proposed framework}
During the learning phase, we first use the $K$-means algorithm to cluster the training set into $K$ groups, and then, an ADC-cycleGAN model is trained using the samples of each group, i.e., totally $K$ ADC-cycleGAN models is trained. During the test phase, each test sample is clustered based on the centroids of the $k$-means algorithm obtained by the training set, and then, the corresponding ADC-cycleGAN with respect to the cluster number is used to synthesize image.

\section{Experimental results}
\label{Sec:exp}
This section conducts a number of experiments to evaluate the effectiveness of ADC-cycleGAN in synthesizing MR images from CT scans and vice versa from unpaired data and compare it with other medical image synthesis methods such as CycleGAN~\cite{zhu2017unpaired}, NiceGAN~\cite{chen2020reusing}, UGATIT~\cite{citation-0}, RegGAN~\cite{kong2021breaking} and DC-cycleGAN~\cite{wang2022dc}. Their codes are obtained from their official GitHub~\footnote{https://github.com/simontomaskarlsson/CycleGAN-Keras}
\footnote{https://github.com/alpc91/NICE-GAN-pytorch} \footnote{https://github.com/taki0112/UGATIT} 
\footnote{https://github.com/Kid-Liet/Reg-GAN}
\footnote{https://github.com/JiayuanWang-JW/DC-cycleGAN}
In this study, CycleGAN is used as the baseline model, because it has been widely used in generating medical images from unpaired data.\par


\subsection{Evaluation indexes}
\label{Sec:sec:metrics}
We used three evaluation metrics, namely SSIM~\cite{Zhou2004Image}, MAE, and PSNR for performance evaluation and compression.\par

\textbf{SSIM} is a common evaluation metric that measures the similarity between two images ($x_1$ and $x_2$) by considering luminance $\emph{l}$, contrast $\emph{c}$, and structure $\emph{s}$. Eq.~(\ref{SSIM}) can be used to compute SSIM. 

\textbf{MAE} measures the average absolute between two images. In other words, MAE computes the distance between real and synthesized images. It can be written as:  
\begin{align}
MAE=\frac{1}{H W} \sum _{i=1}^{H} \sum _{j=1}^{W}\vert x_{1}(i, j)-x_{2}(i, j)\vert,     
\end{align}
where $H$ and $W$ are the high and wide of the images, respectively.

\textbf{PSNR} is another indicator that can be used to assess the quality of the synthesized image, which can be computed as:
\begin{align}
PSNR=10 \log _{10}(L / \mathrm{MSE}),
\end{align}
where $L$ is the dynamic range of the pixel values, and:  
\begin{align}
MSE=\frac{1}{HW} \sum_{i=1}^{H} \sum_{j=1}^{W}(x_{1}(i, j)-x_{2}(i, j))^{2}.
\end{align}

A large PSNR indicates that the two images are closer.\par


\subsection{Dataset}
\label{Sec:sec:Dataset}
Although the aim is medical image synthesis from unpaired data, it is required to use paired data for computing the evaluation metrics, i.e., SSIM, MAE and PSNR. To achieve this, the dataset introduced by Han et al.~\cite{han2017mr} is obtained from their project website\footnote{https://github.com/ChengBinJin/MRI-to-CT-DCNN-TensorFlow}. This dataset includes 367 paired CT and MR images from multiple slices, and the size of each image is 512 $\times$ 256. CT scans have some head frames due to Gamma Knife treatment. {\color{black} As the publicly available dataset we utilized only included images in png format, we were unable to follow the approach used by Han et al.~\cite{han2017mr} for calculating masks based on Hounsfield units and removing head frames. Instead, we manually removed the head frames by cropping the CT scans.}

\subsection{Parameters setting}
\label{Sec:sec:Implementation setting}
ADC-cycleGAN has a similar structure to CycleGAN~\cite{zhu2017unpaired}. It consists of two generators $G$ and $F$ and two discriminators $D_{x}$ and $D_{y}$.
We followed the CycleGAN and set $\lambda$, batch size and the number of epochs to 10, 1 and 200, respectively. The number of clusters of $K$-means algorithm based on the ablation study is set to 4. {\color{black} To have a fair comparison, we follow our previous work for $\beta$= 0.5. This value is obtained based on a sensitivity analysis.}
During the training phase, generators are updated five times followed by updating discriminators once.\par


We normalized all images in the range of -1 to 1 and resized them to 256$\times$256. In addition, 90\% and 10\% of the dataset are used for training and testing, respectively. 
In order to have a fair comparison, the $K$-means algorithm is used for all methods, i.e., for each method, we first used $K$-means algorithm to cluster training set into $K$ groups, and then, each group is trained by a model. This procedure is repeated five times for each comparison method and  used the mean value along with the standard deviation (SD) as the final result. 
The experiments were carried out on a server equipped with an Intel(R) Xeon(R) E5-2650 CPU and an Nvidia GTX 1080TI GPU.\par
\subsection{Ablation study}
\label{Sec:sec:Ablation study}
In this section, we conduct ablation studies to show the effects of using different components of the proposed ADC-cycleGAN model.
Two experiments are conducted. The first experiment studies the effect of a different number of clusters $K$ and finds the best value, As such, $K$ is varied from 2 to 5. Tables~\ref{CT-MRI ablation cluster} and \ref{MRI-CT ablation cluster} show the results of ADC-cycleGAN with different number of {\color{black}clusters} for MR and CT synthesis, respectively. ADC-cycleGAN with $K$=4 outperforms other cluster numbers. Therefore, for the rest of experiments $K$ is set to 4.\par
\begin{table}[h]
\centering
\caption{The ablation results for MR synthesis. ``Mean (standard deviation)" for the different number of clusters.}
\begin{adjustbox} {width=\columnwidth}
\label{CT-MRI ablation cluster}
\begin{tabular}{l c c c c}
  \toprule
  \multirow{1}*{\# of clusters} & \multicolumn{1}{c}{MAE $\downarrow$} & \multicolumn{1}{c}{PSNR $\uparrow$} & \multicolumn{1}{c}{SSIM $\uparrow$}\\
  
  \midrule
        2 & 0.13360 (0.00783) & 18.32465 (0.53110) & 0.59565 (0.01087)  \\ 
         \midrule
        3 & 0.12458 (0.00995) & 18.81969 (0.72182) & 0.62591 (0.01589)  \\ 
         \midrule
        4 & \textbf{0.11566} (0.00746) & 19.69240 (0.75123) & \textbf{0.64330} (0.01975)  \\ 
         \midrule
        5 & 0.11897 (0.01354) & \textbf{19.74868} (1.37066) & 0.63929 (0.03150)  \\ 
  \bottomrule
\end{tabular}
\end{adjustbox}
\end{table}

\begin{table}[h]
\centering
\caption{The ablation results for CT synthesis. ``Mean (standard deviation)" for the different number of clusters.}
\begin{adjustbox} {width=\columnwidth}
\label{MRI-CT ablation cluster}
\begin{tabular}{l c c c c}
  \toprule
  \multirow{1}*{\# of clusters} & \multicolumn{1}{c}{MAE $\downarrow$} & \multicolumn{1}{c}{PSNR $\uparrow$} & \multicolumn{1}{c}{SSIM $\uparrow$}\\
  
  \midrule
        2 & 0.12722 (0.00632) & 16.74714 (0.81303) & 0.69372 (0.01155)  \\ 
         \midrule
        3 & 0.11166 (0.00571) & 17.58067 (0.60216) & 0.71480 (0.01184)  \\ 
         \midrule
        4 & \textbf{0.10809} (0.00292) & \textbf{17.95930} (0.52605) & \textbf{0.71589} (0.00655)  \\ 
         \midrule
        5 & 0.12328 (0.01482) & 17.52259 (1.75595) & 0.69916 (0.02603)  \\ 
  \bottomrule
\end{tabular}
\end{adjustbox}
\end{table}

The second experiment shows the impact of using DC loss, attention mechanism and clustering method in the ADC-cycleGAN structure. Four different combinations are used, as follows: 
\begin{itemize}
\item CycleGAN (wo): the baseline model.
\item CycleGAN (w): CycleGAN with clustering technique.
\item A-cycleGAN (wo): CycleGAN with CBAM.
{\color{black}
\item A-cycleGAN (w): CycleGAN with CBAM and clustering technique.}
\item DC-cycleGAN (wo): CycleGAN with DC loss. 
\item DC-cycleGAN (w): CycleGAN with DC loss and clustering technique.
\item ADC-cycleGAN (wo): CycleGAN with DC loss and CBAM.
\item ADC-cycleGAN (w): CycleGAN with DC loss, CBAM and clustering technique.
\end{itemize}

\begin{table}[h]
\centering
\caption{ The ablation study for CT and MR synthesis. The mean (standard deviation) for different conditions. The notions ``(w)'' and ``(wo)'' indicate with or without $K$-means algorithm.}
\begin{adjustbox} {width=\columnwidth}
\label{MRI-CT ablation}
\begin{tabular}{l c c c c}
  \toprule
  \multirow{1}*{Method} & \multicolumn{1}{c}{MAE $\downarrow$} & \multicolumn{1}{c}{PSNR $\uparrow$} & \multicolumn{1}{c}{SSIM $\uparrow$}\\
  
  \midrule
        CycleGAN (wo) & 0.13789 (0.01073) & 16.47120 (0.57922) & 0.64637 (0.00892)  \\
        \midrule
        CycleGAN (w) & 0.12245 (0.01008) & 17.70577 (0.79366) & 0.66171 (0.01509)  \\
        \midrule
        A-cycleGAN (wo) & 0.14080 (0.00989)	& 16.28680 (0.55052) & 0.64795 (0.00870)\\
        \midrule
        A-cycleGAN (w) & 0.12537 (0.00847) & 17.39279 (0.52072) & 0.66055 (0.01228)  \\ 
        \midrule
        DC-cycleGAN (wo) & 0.14157 (0.00646) & 16.30498 (0.36758) & 0.63576 (0.00617)  \\
        \midrule
        DC-cycleGAN (w) & 0.11069 (0.00401) & 18.80157 (0.35545) & 0.67907 (0.00947)  \\ 
        \midrule
        ADC-cycleGAN (wo) & 0.13622 (0.00235) & 16.57089 (0.11158) & 0.64359 (0.00412)  \\ 
        \midrule
        ADC-cycleGAN (w) & \textbf{0.11005} (0.00450) & \textbf{19.04385} (0.48771) & \textbf{0.68551} (0.00849)  \\ 
  \bottomrule
\end{tabular}
\end{adjustbox}
\end{table}

{\color{black}Table~\ref{MRI-CT ablation} presents the results of the second ablation study. As can be seen, ADC-cycleGAN outperforms all other combinations. This finding confirms that every component plays a crucial role in synthesizing both MR and CT images.
Moreover, our proposed methods with attention mechanisms, i.e., ADC-cycleGAN (wo) and ADC-cycleGAN (w), performed better and more stable results as compared with those without attention mechanisms, i.e., DC-cycleGAN (wo) and DC-cycleGAN (w). However, the baseline models, i.e., CycleGAN (wo) and CycleGAN (w), perform slightly better than those with attention mechanisms, i.e., A-CycleGAN (wo) and A-CycleGAN (w).  
In addition, all methods that use the clustering mechanism, i.e., {\color{black} CycleGAN(w), A-cycleGAN(w), DC-cycleGAN(w), and ADC-cycleGAN(w)} outperform those methods without the clustering algorithm, {\color{black}i.e., CycleGAN(wo), A-cycleGAN(wo), DC-cycleGAN(wo), and ADC-cycleGAN(wo).} This proves the capability of the $K$-means algorithm in synthesizing high-quality images from datasets that contain images with various structures. However, the $K$-means algorithm reduces the stability of the model.  }

\subsection{Comparison with other methods}
\label{Sec:sec:comp}
This section compares the performance of ADC-cycleGAN with CycleGAN~\cite{zhu2017unpaired}, NiceGAN~\cite{chen2020reusing}, UGATIT~\cite{citation-0}, RegGAN~\cite{kong2021breaking} and DC-cycleGAN~\cite{wang2022dc}. For all methods, the $K$-means clustering algorithm (with $K=4$) is used to group the training samples into $K$ clusters, and then, each group is trained by a model, i.e., $K=4$ models are trained for each method. Note that, during the test phase, the centroids of the $K$-means algorithm obtained during training are used to cluster the test set. Then, for each test sample, the trained model with respect to its cluster is used to generate the image.\par 

\begin{table}[h]
\centering
\caption {The mean (standard deviation) values of the MR synthesis quality evaluation metrics for different methods.}
\begin{adjustbox} {width=\columnwidth}
\label{CT-MRI result}
\begin{tabular}{l c c c c}
  \toprule
  \multirow{1}*{Method} & \multicolumn{1}{c}{MAE $\downarrow$} & \multicolumn{1}{c}{PSNR $\uparrow$} & \multicolumn{1}{c}{SSIM $\uparrow$}\\
  
  \midrule
        CycleGAN & 0.12636 (0.01343) & 18.52153 (1.06237) & 0.62686 (0.02044)\\ 
         \midrule
        NiceGAN & 0.12562 (0.00145) & 18.34712 (0.12351) & 0.62077 (0.00280)\\ 
         \midrule
        UGATIT & 0.10647 (0.00320) & 19.92874 (0.20880) & 0.64863 (0.00463)\\
         \midrule
        RegGAN & \textbf{0.09704} (0.00162) & \textbf{20.45722} (0.18531) & \textbf{0.66536} (0.00191)\\ 
         \midrule
        DC-cycleGAN & 0.11395 (0.00406) & 19.85754 (0.38057) & 0.64502 (0.01302)\\ 
         \midrule
        ADC-cycleGAN & 0.11080 (0.00571) & 20.12068 (0.55074) & 0.65568 (0.01215)\\
  \bottomrule
\end{tabular}
\end{adjustbox}
\end{table}

\begin{table}[h]
\centering
\caption{ The mean (standard deviation) values of the CT synthesis quality evaluation metrics for different methods.}
\begin{adjustbox} {width=\columnwidth}
\label{MRI-CT result}
\begin{tabular}{l c c c c}
  \toprule
  \multirow{1}*{Method} & \multicolumn{1}{c}{MAE $\downarrow$} & \multicolumn{1}{c}{PSNR $\uparrow$} & \multicolumn{1}{c}{SSIM $\uparrow$}\\
  
  \midrule
        CycleGAN & 0.11853 (0.00673) & 16.89000 (0.52494) & 0.69656 (0.00973)  \\
         \midrule
        NiceGAN & 0.12742 (0.00120) & 15.99856 (0.15009) & 0.68270 (0.00175)  \\ 
         \midrule
        UGATIT & 0.10825 (0.00328) & 17.04600 (0.18058) & 0.70392 (0.00612) \\
         \midrule
        RegGAN & 0.13940 (0.04498) & 16.21307 (1.03060) & 0.67735 (0.02595)  \\
         \midrule
        DC-cycleGAN & \textbf{0.10742} (0.00395) & 17.74560 (0.33032) & 0.71312 (0.00592) \\
         \midrule
        ADC-cycleGAN & 0.10931 (0.00329) & \textbf{17.96701} (0.42467) & \textbf{0.71534} (0.00482) \\
  \bottomrule
\end{tabular}
\end{adjustbox}
\end{table}

\begin{table}[h]
\centering
\caption{The results (Mean (standard deviation)) of bidirectional MR and CT synthesis.}
\begin{adjustbox} {width=\columnwidth}
\label{average result}
\begin{tabular}{l c c c c}
  \toprule
  \multirow{1}*{Method} & \multicolumn{1}{c}{MAE $\downarrow$} & \multicolumn{1}{c}{PSNR $\uparrow$} & \multicolumn{1}{c}{SSIM $\uparrow$}\\
  
  \midrule
        CycleGAN & 0.12244 (0.01008) & 17.70576 (0.79366) & 0.66171 (0.01508)  \\ 
         \midrule
        NiceGAN & 0.12652 (0.00132) & 17.17284 (0.13680) & 0.65174 (0.00228)  \\ 
         \midrule
        UGATIT & \textbf{0.10736} (0.00324) & 18.48737 (0.19469) & 0.67628 (0.00537)  \\ 
         \midrule
        RegGAN & 0.11822 (0.02330) & 18.33515 (0.60796) & 0.67136 (0.01393)  \\ 
         \midrule
        DC-cycleGAN & 0.11069 (0.00401) & 18.80157 (0.35545) & 0.67907 (0.00947)  \\ 
         \midrule
        {\color{black}ADC-cycleGAN} & 0.11005 (0.00450) & \textbf{19.04385} (0.48771) & \textbf{0.68551} (0.00849)  \\ 
  \bottomrule
\end{tabular}
\end{adjustbox}
\end{table}

The quantitative results, i.e., MAE, PSNR, and SSIM, for MR and CT synthesis are shown in Tables~\ref{CT-MRI result} and \ref{MRI-CT result}, respectively. RegGAN produces the best results in synthesizing MR images from CT scans, and ADC-cycleGAN is ranked as the second-best method in terms of PSNR and SSIM, and the third method in term of MAE. In contrast, ADC-cycleGAN outperforms all methods in terms of PSNR and SSIM in generating CT scans from MR images, while DC-cycleGAN is ranked as the first method in term of MAE. However, RegGAN produces inferior results in synthesizing CT scans from MR images and it is not able to perform bidirectional learning, i.e., it is required to train two times, one for CT-to-MR and one for MR-to-CT.
Overall, our proposed ADC-cycleGAN model is able to produce the best results in generating both MR and CT images, i.e., bidirectional learning, in terms of PSNR and SSIM, while UGATIT and ADC-cycleGAN perform similarly in terms of MAE (see Table~\ref{average result}).
{\color{black}While our method can produce comparable results as compared with other models, there are several limitations that must be addressed.
First, The effectiveness of ADC-cycleGAN and other methods is largely influenced by the quality of the training dataset. When there exist artifacts on the head frames of CT scans, all methods produce inferior results. Second, our ADC-cysleGAM model still suffers from the model collapse issue that causes by the presence of various slices in the dataset. Although we mitigated this issue by employing the $K$-mean clustering algorithm, further research is required to find more effective solutions.\par }


\begin{figure}[h]
	\centering
	\subfloat[]{
		\begin{minipage}[t]{0.13\textwidth}
			\centering
			\includegraphics[width=0.95\textwidth]{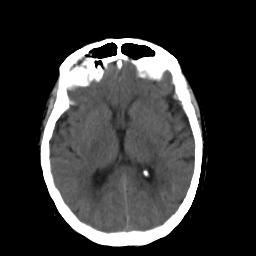}\\
			\vspace{0.1cm}
			\includegraphics[width=0.95\textwidth]{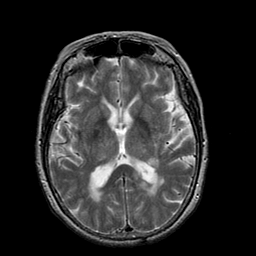}\\
			\vspace{0.3cm}
		\end{minipage}%
	}%
	\subfloat[]{
		\begin{minipage}[t]{0.13\textwidth}
			\centering
			\includegraphics[width=0.95\textwidth]{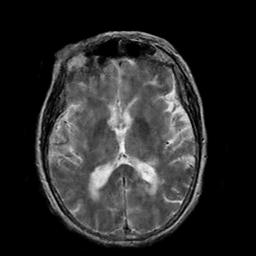}\\
			\vspace{0.1cm}
			\includegraphics[width=\textwidth]{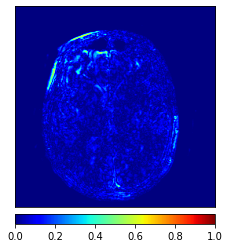}\\
			\vspace{0.1cm}
		\end{minipage}%
	}%
	\subfloat[]{
		\begin{minipage}[t]{0.13\textwidth}
			\centering
			\includegraphics[width=0.95\textwidth]{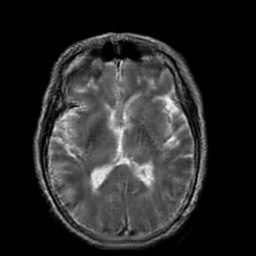}\\
			\vspace{0.1cm}
			\includegraphics[width=\textwidth]{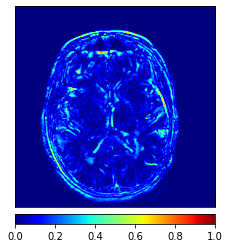}\\
			\vspace{0.1cm}
		\end{minipage}%
	}%
	\subfloat[]{
		\begin{minipage}[t]{0.13\textwidth}
			\centering
			\includegraphics[width=0.95\textwidth]{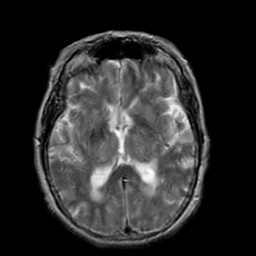}\\
			\vspace{0.1cm}
			\includegraphics[width=\textwidth]{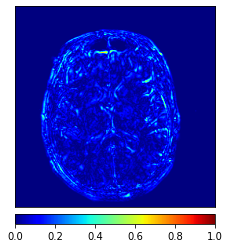}\\
			\vspace{0.1cm}
		\end{minipage}%
	}%
	\subfloat[]{
		\begin{minipage}[t]{0.13\textwidth}
			\centering
			\includegraphics[width=0.95\textwidth]{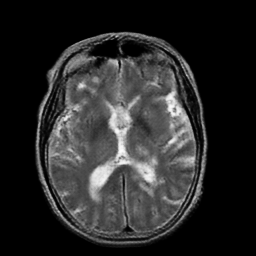}\\
			\vspace{0.1cm}
			\includegraphics[width=\textwidth]{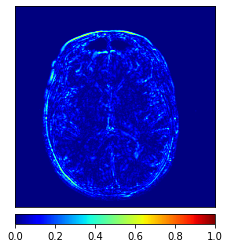}\\
			\vspace{0.1cm}
		\end{minipage}%
	}%
	\subfloat[]{
		\begin{minipage}[t]{0.13\textwidth}
			\centering
			\includegraphics[width=0.95\textwidth]{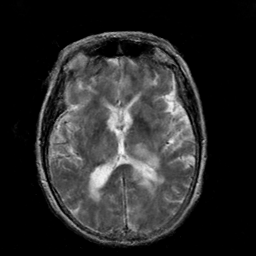}\\
			\vspace{0.1cm}
			\includegraphics[width=\textwidth]{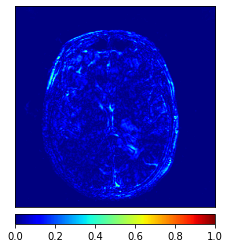}\\
			\vspace{0.1cm}
		\end{minipage}%
	}%
	\subfloat[]{
		\begin{minipage}[t]{0.13\textwidth}
			\centering
            \includegraphics[width=0.95\textwidth]{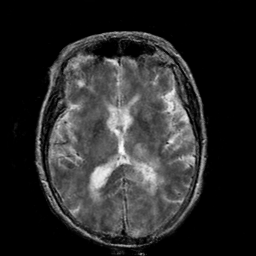}\\
			\vspace{0.1cm}
			\includegraphics[width=\textwidth]{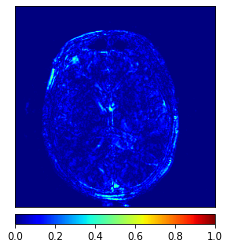}\\
			\vspace{0.1cm}
		\end{minipage}%
	}%
	
	\centering
	\caption{Synthesized MR images along with absolute error maps between groundtruth and synthesized images by different methods. (a) Real image, (b) CycleGAN, (c) NiceGAN, (d) UGATIT, (e) RegGAN, (f) DC-cycleGAN, (g) ADC-cycleGAN}
	\vspace{-1cm}
	\label{vision result MR}
\end{figure}

\begin{figure}[h]
	\centering
	\subfloat[]{
		\begin{minipage}[t]{0.13\textwidth}
			\centering
			\includegraphics[width=0.95\textwidth]{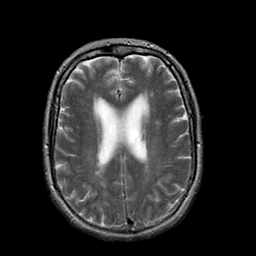}\\
			\vspace{0.1cm}
			\includegraphics[width=0.95\textwidth]{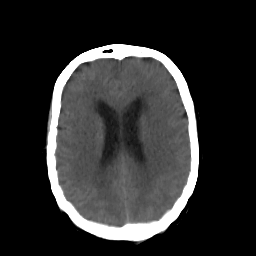}\\
			\vspace{0.3cm}
		\end{minipage}%
	}%
	\subfloat[]{
		\begin{minipage}[t]{0.13\textwidth}
			\centering
			\includegraphics[width=0.95\textwidth]{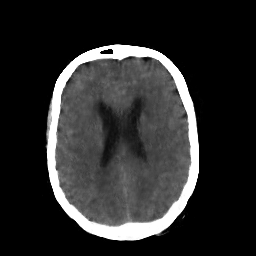}\\
			\vspace{0.1cm}
			\includegraphics[width=\textwidth]{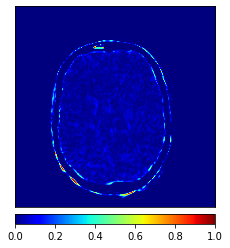}\\
			\vspace{0.1cm}
		\end{minipage}%
	}%
	\subfloat[]{
		\begin{minipage}[t]{0.13\textwidth}
			\centering
			\includegraphics[width=0.95\textwidth]{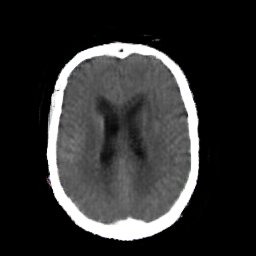}\\
			\vspace{0.1cm}
			\includegraphics[width=\textwidth]{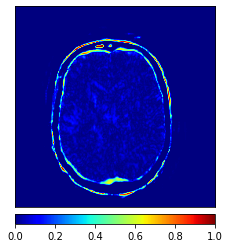}\\
			\vspace{0.1cm}
		\end{minipage}%
	}%
	\subfloat[]{
		\begin{minipage}[t]{0.13\textwidth}
			\centering
			\includegraphics[width=0.95\textwidth]{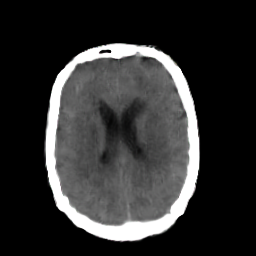}\\
			\vspace{0.1cm}
			\includegraphics[width=\textwidth]{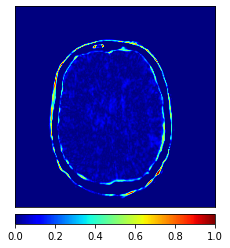}\\
			\vspace{0.1cm}
		\end{minipage}%
	}%
	\subfloat[]{
		\begin{minipage}[t]{0.13\textwidth}
			\centering
			\includegraphics[width=0.95\textwidth]{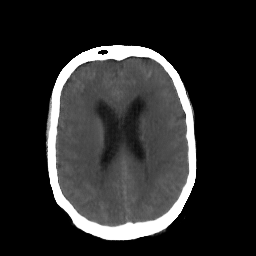}\\
			\vspace{0.1cm}
			\includegraphics[width=\textwidth]{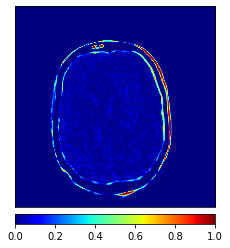}\\
			\vspace{0.1cm}
		\end{minipage}%
	}%
	\subfloat[]{
		\begin{minipage}[t]{0.13\textwidth}
			\centering
			\includegraphics[width=0.95\textwidth]{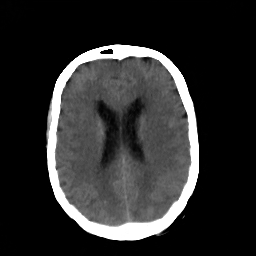}\\
			\vspace{0.1cm}
			\includegraphics[width=\textwidth]{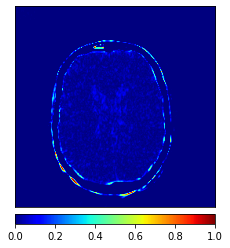}\\
			\vspace{0.1cm}
		\end{minipage}%
	}%
	\subfloat[]{
		\begin{minipage}[t]{0.13\textwidth}
			\centering
			\includegraphics[width=0.95\textwidth]{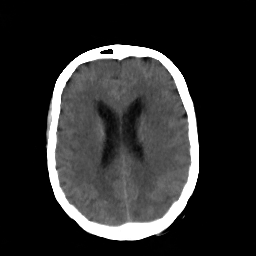}\\
			\vspace{0.1cm}
			\includegraphics[width=\textwidth]{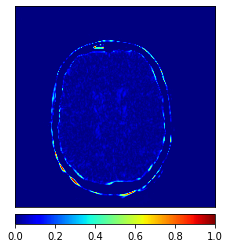}\\
			\vspace{0.1cm}
		\end{minipage}%
	}%
	
	\centering
	\caption{Synthesized CT images and their corresponding absolute error maps for different methods. (a) Real image, (b) CycleGAN, (c) NiceGAN, (d) UGATIT, (e) RegGAN, (f) DC-cycleGAN, (g) ADC-cycleGAN}
	\vspace{-0.5cm}
	\label{vision result CT}
\end{figure}

In addition, the synthesized CT images and their corresponding absolute error maps for different methods are shown in Figs.~\ref{vision result MR} and \ref{vision result CT}, respectively. 
{\color{black}As compared with other methods, the synthesized MR and CT images by ADC-cycleGAN demonstrate a higher level of fidelity to the real images. In particular, it can effectively capture the intricate details in soft tissue within the synthesized MR scans, and accurately replicate the edge structures in the synthesized CT scans. These results indicate the effectiveness of the proposed ADC-cycleGAN in synthesizing MR and CT images.}


\section{Conclusion}
\label{Sec:con}
{\color{black}
In this study, we proposed a bidirectional generative model based on CycleGAN with an attention mechanism for synthesizing medical images from unpaired data. We introduced a dual contrast loss that leverages samples from the source domain as negative samples, pushing the synthesized images further away from the source domain. To capture important features in both channel and space domains, we integrated CBAM into the generators. Additionally, to enable the model to synthesize high-quality images from datasets with varying slice numbers, we employed the $K$-means algorithm to cluster the training set into groups and trained a model for each group. The experimental results, along with the ablation study, demonstrate the effectiveness of the proposed method in synthesizing MR from CT scans and vice versa.\par

In our future research, we plan to improve the quality of synthesized images by developing novel structures for generators and discriminators, aiming to enhance model performance and stability. Additionally, we aim to utilize the synthesized images for solving other tasks such as segmentation. Furthermore, we plan to develop a new dataset for performance evaluation.}


\section*{Declarations}

The dataset analyzed during the current study are available in the Github repository, https://github.com/ChengBinJin/MRI-to-CT-DCNN-TensorFlow.

\bibliography{ref}


\end{document}